\documentclass[english,aps,prl,twocolumn,nopacs,groupedaddress,floatfix]{revtex4}

\usepackage{amsmath,amsfonts,amssymb,graphics,graphicx,epsfig,color,times,bbm}

\newcommand{\id}{\mathbbm{1}}

\newtheorem{theorem}{Theorem}
\newtheorem{lemma}{Lemma}

\begin{document}

\title{Gently modulating opto-mechanical systems}

\author{A.\ Mari and J.\ Eisert}

\affiliation{Institute of Physics and Astronomy, University of Potsdam, 
D-14476 Potsdam, Germany}
\affiliation{Institute for Advanced Study Berlin, D-14193 Berlin, Germany}

\begin{abstract}
We introduce a framework of 
opto-mechanical systems that are 
driven with a mildly amplitude-modulated 
light field, but that are not subject to 
classical feedback or squeezed input light. 
We find that in such a system one can achieve large degrees of 
squeezing of a mechanical micromirror --
signifying quantum properties of opto-mechanical systems --
without the need of any feedback and control, and within 
parameters reasonable in experimental settings. Entanglement dynamics is
shown of states following classical quasi-periodic
orbits in their first moments. 
We discuss the complex time-dependence  
of the modes of a cavity-light field
and a mechanical mode in phase space. Such settings
give rise to certifiable quantum properties within experimental conditions
feasible with present technology. 
\end{abstract}

\maketitle

Periodically driven quantum systems exhibit a rich behavior
and display non-equilibrium properties that are absent in their static counterparts. 
By appropriately exploiting time-periodic
driving, strongly correlated Bose-Hubbard-type models 
can be dynamically driven to quantum phase transitions
\cite{Holthaus},
systems can be dynamically decoupled from their
environments 
to avoid decoherence in quantum
information science \cite{Decoupling}, 
and quite intriguing dynamics of Rydberg atoms strongly 
driven by microwaves \cite{Rydberg} can arise. 
It has also been muted that such
time-dependent settings may give rise to 
entanglement dynamics in oscillating molecules \cite{Briegel}. A framework of such 
periodically driven systems is provided by the theory of linear differential
equations with periodic coefficients or 
inhomogeneities, including Floquet's theorem \cite{Floquet}.

In this work, we aim at transferring
such ideas to describe a new and in fact
quite simple regime of opto-mechanical systems, of micromirrors as part of
a Fabry-Perot cavity \cite{Metzker,Aspe,Schliesser,Groeblacher}:
So to one of the settings \cite{LaHaye,Vitali,VedralVitali,Rabl,Review}
that are the most promising candidates in the race of 
exploring certifiable quantum effects involving macroscopic
mechanical modes. This is an instance of a 
regime of driving with mildly amplitude-modulated light.
We find that in this regime, high 
degrees of squeezing below the vacuum noise level
can be reached, signifying genuine quantum dynamics. 
More specifically, in contrast to earlier descriptions of 
opto-mechanical systems with a periodic time-dependence in 
some aspect of the description, we will not rely on classical 
feedback based on processing of measurement-outcomes -- a 
promising idea in its own right in a continuous-measurement
perspective \cite{Clerk,Woolley} -- 
or resort to driving with squeezed light. Instead,
we will consider the plain setting of a time-periodic amplitude
modulation of an input light. The picture developed here, based
in the theory of differential equations,
gives rise to a framework of describing such situations. We find
that large degrees of squeezing can be reached (complementing
other very recent non-periodic approaches based on cavity-assisted
squeezing using an additional squeezed light beam \cite{Jaehne}).
It is the practical appeal of this work that such quantum
signatures can be reached without the necessity of
any feedback, no driving with additional fields, and no 
squeezed light input (the scheme by far outperforms 
direct driving with a single squeezed light mode): 
In a nutshell, one has to 
simply gently shake the
system in time with the right frequency to have the mechanical
and optical modes rotate appropriately around each other, reminding
of parametric amplification, and
to so directly certify quantum features of such a system. 

{\it Time-dependent picture of system.}
Before we discuss the actual time-dependence of the
driven system, setting the stage, we start our description
with the familiar Hamiltonian of a
system of a Fabry-Perot cavity of length $L$ and finesse $F$ being 
formed on one end by a moving micromirror,\begin{eqnarray}
	&& H=\hbar\omega_{c}a^{\dagger}a+\frac{1}{2}\hbar\omega_{m}(p^{2}+q^{2})-
	\hbar G_{0}a^{\dagger}a q \nonumber  \\ 
	&& +i\hbar \sum_{n=-\infty}^{\infty} ( E_n e^{-i(\omega_{0}+n 
	\Omega)t}a^{\dagger}- E_n^* e^{i(\omega_{0}+n \Omega)t}a).  
	\label{ham0}
\end{eqnarray}
Here, $\omega_m$ is the 
frequency of the mechanical mode with quadratures $q$ and $p$
satisfying the usual commutation relations of 
canonical coordinates, while the bosonic operators $a$ and $a^\dag$ 
are associated to the cavity mode with frequency $\omega_c$ and decay 
rate $\kappa=\pi c / (2 F L)$. 
$G_0=\sqrt{\hbar/(m \omega_m)}\, \omega_c/L $ is the coupling coefficient 
of the radiation pressure, where $m$ is the effective mass of the mode of the mirror being used. 
Importantly, we allow for any periodically modulated driving at this point, which can be 
expressed in such a Fourier series, where $\Omega= 2 \pi / \tau$ and $\tau>0$ is the modulation period.
The main frequency of the driving field is $\omega_0$ while the modulation 
coefficients $\{E_n\}$ are related to the power of the associated sidebands 
$\{P_n\}$ by $|E_n|^2=2 \kappa P_n/(\hbar \omega_0)$.
The resulting dynamics under this Hamiltonian 
together with an unavoidable 
coupling of the mechanical mode to a thermal reservoir and cavity losses
gives rise
to the quantum Langevin equation in the 
reference frame rotating with frequency $\omega_0$, $\dot{q}=\omega_m p$, and
\begin{eqnarray}\label{QLE}
\dot{p}&=&-\omega_m q - \gamma_m p + G_0 a^{\dag}a + \xi,  \\
\dot{a}&=&-(\kappa+i\Delta_0)a +i G_0 a q +\sum_{n=-\infty}^{\infty} E_n e^{-in \Omega t} +\sqrt{2\kappa} a^{in}, \nonumber
\end{eqnarray}
where $\Delta_0=\omega_c-\omega_0$ is the cavity detuning. $\gamma_m$ is here an
effective damping rate related to the oscillator quality factor $Q$ by $\gamma_m=\omega_m/Q$. The mechanical $(\xi)$ and the optical $(a^{in})$ noise operators have zero mean values and are characterized by their auto correlation functions which, in the Markovian approximation, are
\begin{eqnarray} \label{Marknoise}
	 \langle \xi(t) \xi (t')+ \xi(t') \xi (t) \rangle/2&=&\gamma_m (2 \bar n +1) \delta(t-t')
\end{eqnarray}
and 
$\langle a^{in}(t) a^{in \dag}(t') \rangle=\delta(t-t')$, where $\bar n=[\exp(\frac{\hbar \omega_m}{k_B T})-1]^{-1}$ is the mean thermal phonon number.
Here, we have implicitly assumed that such an effective damping model holds \cite{Pedantic}, 
which is a reasonable assumption in a wide range of parameters 
including the current experimental regime.

{\it Semiclassical phase space orbits.}
Our strategy of a solution will be as follows: we will first 
investigate the classical phase space orbits of the first moments of quadratures. 
We then consider the quantum fluctuations around 
the asymptotic quasi-periodic orbits, 
by implementing the usual linearization of the Heisenberg 
equations of motion \cite{Vitali,VedralVitali}
(excluding the very weak driving regime). Exploiting 
results from the theory of linear 
differential equations with periodic coefficients, we can then proceed to 
describe the dynamics of fluctuations and find an analytical solution 
for the second moments.

If we average the Langevin 
equations (\ref{QLE}), assuming $\langle a^\dag a\rangle\simeq|\langle  a\rangle|^2$, 
$\langle a q\rangle\simeq
\langle a \rangle\langle q \rangle$
(true in the semi-classical driving regime we are interested in), we have a
nonlinear differential equation for the first moments. 
Far away from instabilities and multi-stabilities, 
a power series ansatz in the coupling $G_0$
$\langle O \rangle(t)=\sum_{j=0}^{\infty} O_j(t) G_0^j$,
is justified, where $O=a,p,q$. 
If we substitute this expression into the averaged Langevin equation (\ref{QLE}), we 
get a set of recursive  differential equation for the variables $O_j(.)$. The only two 
nonlinear terms in Eq.\ (\ref{QLE}) are both proportional to $G_0$, 
therefore, for each $j$, the differential equation for the set of unknown variables $O_j(.)$ 
is a \textit{linear} inhomogeneous system 
with constant coefficients and $\tau$-periodic driving. 
 Then, after an exponentially decaying 
 initial transient (of the order of $1/\gamma_m$), the 
 asymptotic solutions $O_j$ will have the same periodicity of 
 the  modulation \cite{Floquet}, justifying the Fourier expansion
\begin{equation}\label{double}
\langle O \rangle(t)= \sum_{j=0}^{\infty}\sum_{n=-\infty}^{\infty} O_{n,j} e^{i n \Omega t}G_0^j.
\end{equation}
Substituting this in Eq.\ (\ref{QLE}), we find the following 
recursive formulae for the time independent coefficients $O_{n,j}$,
\begin{eqnarray}
	q_{n,0}&=&p_{n,0}=0, \\
	 a_{n,0}&=&{E_{-n}}/({\kappa	
	+i(\Delta_0 + n \Omega)}),
\end{eqnarray}	
corresponding to the zero coupling limit $G_0= 0$, while for each 
$j\ge 1$, we have
\begin{eqnarray}
	q_{n,j}&=&\omega_m \sum_{k=0}^{j-1}\sum_{m=-\infty}^{\infty} 
	\frac{  a_{m,k}^*\; a_{n+m,j-k-1}}{ \omega_m^2-n	\Omega^2+i\gamma_m n \Omega}, \\
	p_{n,j}&=&\frac{i n \Omega}{\omega_m} q_{n,j},\,
	a_{n,j}=i \sum_{k=0}^{j-1}\sum_{m=-\infty}^{\infty} \frac{a_{m,k} 
	q_{n-m,j-k-1}}{\kappa +i (\Delta_0 + n \Omega)},\nonumber
\end{eqnarray} 
Within the typical parameter space, considering only the first terms in the 
double expansion (\ref{double}), corresponding to the first side bands, leads to a good analytical 
approximation of the classical periodic orbits, see Fig.\ 1. 
On physical grounds, this is expected 
to be a good approximation, since $G_0\ll\omega_m$, 
and because high side-bands (of frequency $n \Omega$) fall outside the
cavity bandwidth, $n \Omega>2\kappa$. What is more, 
the decay behavior of $E_n$ related to the smoothness of the drive
inherits a good approximation in terms of few sidebands. 

\begin{figure}[tbh]
	\centerline{\includegraphics[width=0.45\textwidth]{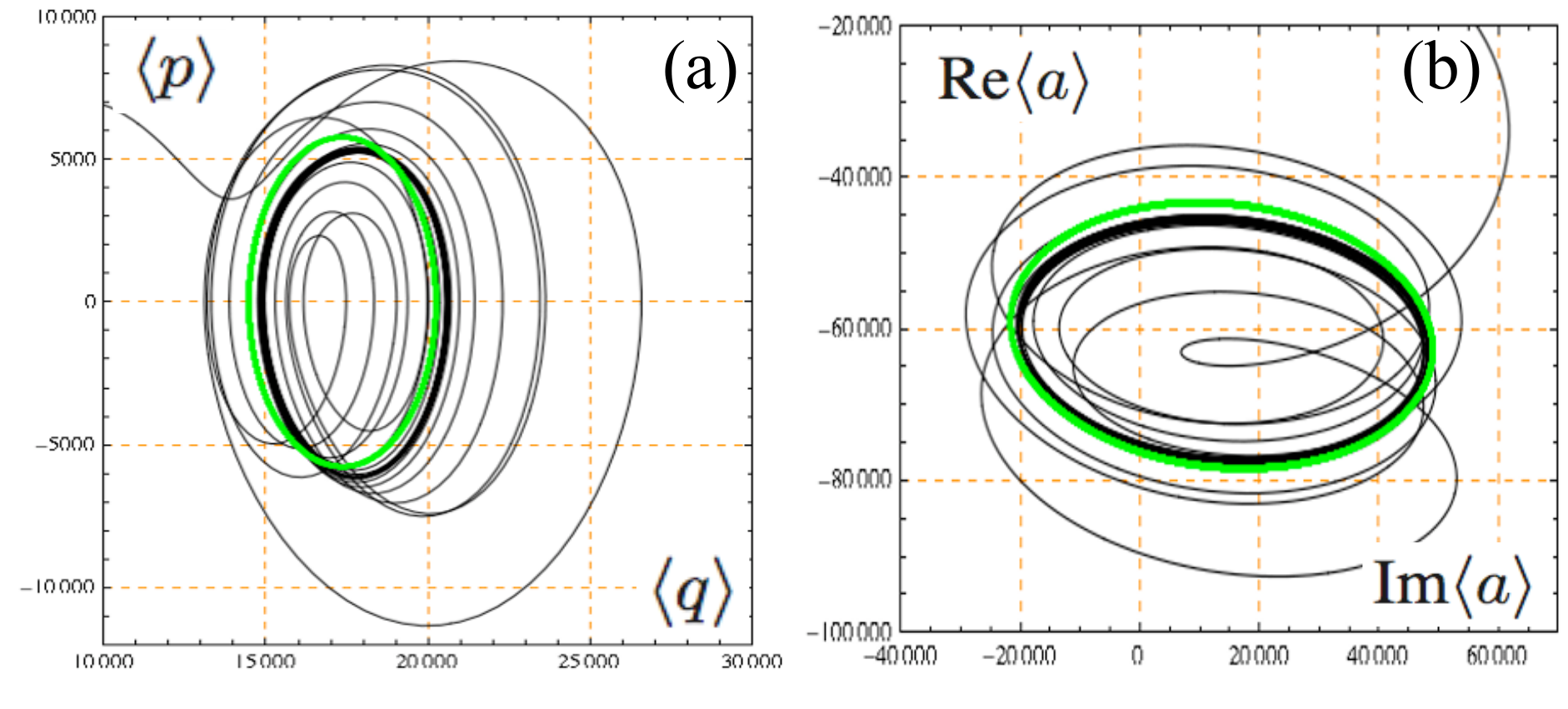}} 
	\caption{Phase space trajectories of the first moments of the mirror (a) 
	and light (b) modes. Numerical simulations for $t \in [0,50\tau]$ 
	(black) and analytical approximations of the 
	asymptotic orbits (green).} \label{orbits}
\end{figure}

{\it Quantum fluctuations around the classical orbits.}
We will now turn to the actual quantum dynamics
taking first moments into account separately
when writing any operator as
$O(t)=\langle O \rangle (t) + \delta O(t)$. The
frame will hence be provided by the motion of the first moments.
In this reference frame, as long as 
$|\langle a \rangle |\gg1$, the usual linearization
approximation to (\ref{QLE}) can be implemented.
In what follows, we will also use the quadratures
$\delta x=(\delta a+\delta a^\dag)/\sqrt{2}$ and $\delta y=-i(\delta a-\delta a^\dag)/\sqrt{2}$, and the 
analogous
input noise quadratures $x^{in}$ and $y^{in}$.
For the vector of all quadratures we will write $u=(\delta q,\delta p,\delta x, \delta y)^T$,
with $n=(0,\xi,\sqrt{2\kappa}x^{in},\sqrt{2\kappa}y^{in})^T$ being the 
noise vector 
  \cite{Vitali,Pedantic}. 
Then the  time-dependent
inhomogeneous equations of motion
arise as $\dot{u}(t)=A(t)u(t)+n(t)$, with 
\begin{eqnarray}\label{LQLE}
  A(t)&=&\left[\begin{array}{cccc}
    	0 	  & \omega_m	& 0 		& 	0 \\
     	-\omega_m & -\gamma_m	& G_x(t)	& G_y(t)  \\
        -G_y(t)	  & 	0	& -\kappa 	& \Delta(t) \\
    	 G_x(t)   &     0	& -\Delta(t) 	& -\kappa
  	\end{array}\right],
\end{eqnarray}
where the real $A(t)$ contains the time-modulated 
coupling constants and the detuning as
$G(t)=G_x(t)+ i G_y(t)$, 
\begin{eqnarray}
G(t)=\sqrt{2}\langle a(t) \rangle G_0,\,\,
\Delta(t)=\Delta_0-G_0 \langle q(t) \rangle.
\end{eqnarray}
From now on we will consider quasi-periodic orbits only, 
so the long-time dynamics following
the initial one, when the first moments follow a 
motion that is $\tau$-periodic. 
Then, $A$ is $\tau$-periodic, and hence
\begin{equation}
	A(t)=A(t+\tau)=\sum_{n=-\infty}^{\infty}A_n e^{i \Omega n t}.
\end{equation}	
In turn, if all eigenvalues of $A(.)$ having negative real parts for all $t\in[0,\tau]$
is a sufficient condition for stability.
From the Markovian assumption (\ref{Marknoise}), we have 
\begin{equation}
	\langle n_i(t) n_j(t')+n_j(t')n_i(t)\rangle/2=\delta(t-t')D_{i,j}, 
\end{equation}	
where 
$D=\text{diag}(0,\gamma_m(2\bar n+1),\kappa,\kappa).$
The formal solution of Eq.\ (\ref{LQLE}) is \cite{Floquet}
\begin{equation}\label{usol}
	u(t)=U(t,t_0)u(t_0)+\int_{t_0}^t U(t,s) n(s)ds,
\end{equation} 
where $U(t,t_0)$ is the principal matrix solution 
of the homogeneous system satisfying 
$\dot U(t,t_0)=A(t)U(t,t_0)$ and $U(t_0,t_0)=\id$.
From Eqs.\ (\ref{LQLE}, \ref{usol}), we have
as an equation of motion of the covariance matrix (CM)
of the 
two modes
\begin{equation}\label{secdiff}
 \dot{V}(t)=A(t)V(t)+V(t)A^T(t)+D.
\end{equation}
Here, the CM $V(.)$
is the $4\times 4$-matrix with components
$V_{i,j}=\langle u_i u_j+u_j u_i\rangle/2$, 
collecting the second 
moments of the quadratures.
This is again an inhomogeneous 
differential equation for the second moments which 
can 
readily be solved using quadrature methods, providing 
numerical solutions that will be used to test and justify analytical approximate
results in important regimes. Moreover, now the 
coefficients and not the inhomogeneity are $\tau$-periodic,
$A(t)=A(t+\tau)$. Again, we can invoke results from the theory of
linear differential equations to Eq.\ (\ref{secdiff})  \cite{Floquet}:
We find that in the long time limit, the CM is periodic 
and can be written as 
\begin{equation}
	V(t)=\sum_n V_n e^{i n\Omega t}.
\end{equation}	
An analytical solution for $V(.)$, 
is most convenient in the Fourier domain, 
$\tilde{f}(\omega)=\int_{-\infty}^{+\infty} e^{-i\omega t}f(t) dt$, giving rise to
\begin{equation}\label{fulle}
	 -i\omega \tilde{u}(\omega)+\sum_{n=-\infty}^{\infty} 
	 A_n \tilde{u}(\omega-n\Omega)=-\tilde{n}(\omega).
\end{equation}
If $A_{n\ne 0}=0$, corresponding to no-modulation, 
we are in the usual regime where the
spectra are centered around $\pm \omega_m$ for the mechanical oscillator 
and around $\pm \Delta$ for the optical mode. The 
modulation introduces sidebands shifted by $\pm n \Omega$. 
If the modulation is weak, 
only the first two sidebands at
$\pm \Omega$ significantly contribute.
For strong modulation also 
further sidebands play a role: Disregarding 
higher sidebands means truncating the summation to $\pm N$, (valid if $u(\omega\pm N \Omega)\simeq 0$). 
Then Eq.\ (\ref{fulle}) can be written as
$\bar A(\omega)\bar u(\omega)=\bar n(\omega)$,
where $\bar u^T(\omega)=(\tilde u^T(\omega-N\Omega),\dots,\tilde u^T(\omega),\dots, \tilde u^T(\omega+N\Omega))$ and $\bar n^T(\omega)=(\tilde n^T(\omega-N\Omega),\dots,\tilde n^T(\omega),\dots, \tilde n^T(\omega+N\Omega))$ are $4\times (2N+1)$ vectors,  
while, in terms of $4\times4$ blocks,
\begin{eqnarray}\label{bigmatrix}{\small
  \bar A(\omega)=\left[\begin{array}{ccccc}
B_{-N} & A_{-1}		     &A_{-2}  	  	   &\cdots  & A_{-2N} \\
A_1	  	&B_{-(N-1)}   & A_{-1} 	   	   &	    & \vdots  \\ 
A_2 	        & A_1     	&B_{-(N-2)} 	   &        &         \\
\vdots	        & 	        &			   	   & 	    & 	      \\
\vdots	        &         	&  		  	  	   &	    & \vdots    \\
A_{2N}	        & \cdots        &  		  	       &A_1  & B_{N} \\
  	\end{array}\right] }
\end{eqnarray}
with $B_k = A_0-i(\omega+k\Omega)$.

From the correlation properties of the noise 
vector $n(.)$, we have that
\begin{eqnarray}
\phi_{i,j}
(\omega,\omega')&=&\langle \bar n_i
(\omega) \bar n_j^*(\omega')+\bar n_j^*(\omega') \bar 
n_i(\omega)\rangle/2\nonumber\\
&=& \sum_{n=-2 N}^{ 2N}\delta (\omega-\omega'-n\Omega) D_n, 
\end{eqnarray}
where
$D_0= \text{diag}(D,D,\dots, D)$, then $D_1$ is the matrix
that has $D$ on all first right off diagonal blocks, $D_2$ on the second off diagonals, 
with $D_n$ analogously defined, and
$D_{-n}=D_n^T$. We now 
define the two frequency correlation function as $\bar V_{i,j}(\omega,\omega')=\langle \bar u_i(\omega) \bar u_j^*(\omega')+\bar u_j^*(\omega') \bar u_i(\omega)\rangle/2$.
We have 
$\bar V(\omega,\omega')=\bar A^{-1} (\omega) \phi(\omega,\omega') [\bar A^{-1} (\omega')]^\dag$.
We are interested only on the central $4\times4$ block of $\bar V$
which we call $\tilde V(\omega,\omega')=[\bar V(\omega,\omega')]_4$.
From $\phi(\omega,\omega')$, we find
\begin{equation}
	\tilde V(\omega,\omega')=\sum_{n=-2N}^{2N} \tilde V_n(\omega) \delta (\omega-\omega'-n \Omega),
\end{equation}	
where 
$\tilde V_n(\omega)=[\bar A^{-1} (\omega)D_n [\bar A^{-1} (\omega-n\Omega)]^\dag ]_4$.
This means that the driving modulation correlates different frequencies, 
but only if they are separated by a multiple of the modulation frequency $\Omega$.
By inverse Fourier transforms we recover the time periodic expression for the CM, 
where the all the components $V_n$ are given by the integral of their noise spectra, i.e.,
\begin{equation}
 V_n=\frac{1}{2\pi} \int_{-\infty}^{+\infty}\tilde V_n(\omega) d\omega.
\end{equation}

{\it Squeezing and entanglement modulation.}
We will now see that the mild amplitude-modulated driving in the cooling regime is exactly the 
tool that we need in order to arrive at strong degrees of squeezing, in the absence of feedback
or squeezed light. We will apply the previous general theory to setting of an optomechanical
system that is experimentally feasible with present technology. 
In fact, all values that we assume have been achieved already and reported on in
publications with the exception of assuming a relatively 
good mechanical $Q$-factor. The reasonable set of experimental parameters 
\cite{Groeblacher} that we assume is 
$L=25$mm, $ F= 1.4 \times10^4$, 
$\omega_m=2\pi$MHz, $Q=10^6$, $m=150$ng, 
$T=0.1$K.  We then consider the -- in the meantime well known -- 
self-cooling regime \cite{Aspe} in which a cavity 
eigen-mode is driven with a red detuned laser $\Delta_0\simeq \omega_m$ 
(with wavelength $\lambda=1064$ nm), but we also add a small 
sinusoidal modulation to the input amplitude with a frequency 
$\Omega=2\omega_m$, so twice the mechanical frequency. 
To be more precise we choose the power of the 
carrier component equal to $P_0=10$mW, and the power of the two 
modulation sidebands equal to $P_{\pm1}=2$mW.

We approximate the asymptotic classical mean values in Eq.\ (\ref{double}) by 
truncating the series only to the first terms with indexes $j=0,\dots ,3$ and $n=-1,0,1$, giving rise to a
good approximation. Fig.\ \ref{orbits} shows that, after less than $50$ 
modulation periods, the first moments reach quasi-periodic orbits which are 
well approximated by our analytical results.

\begin{figure}[tbh]
\centerline{\includegraphics[width=0.5\textwidth]{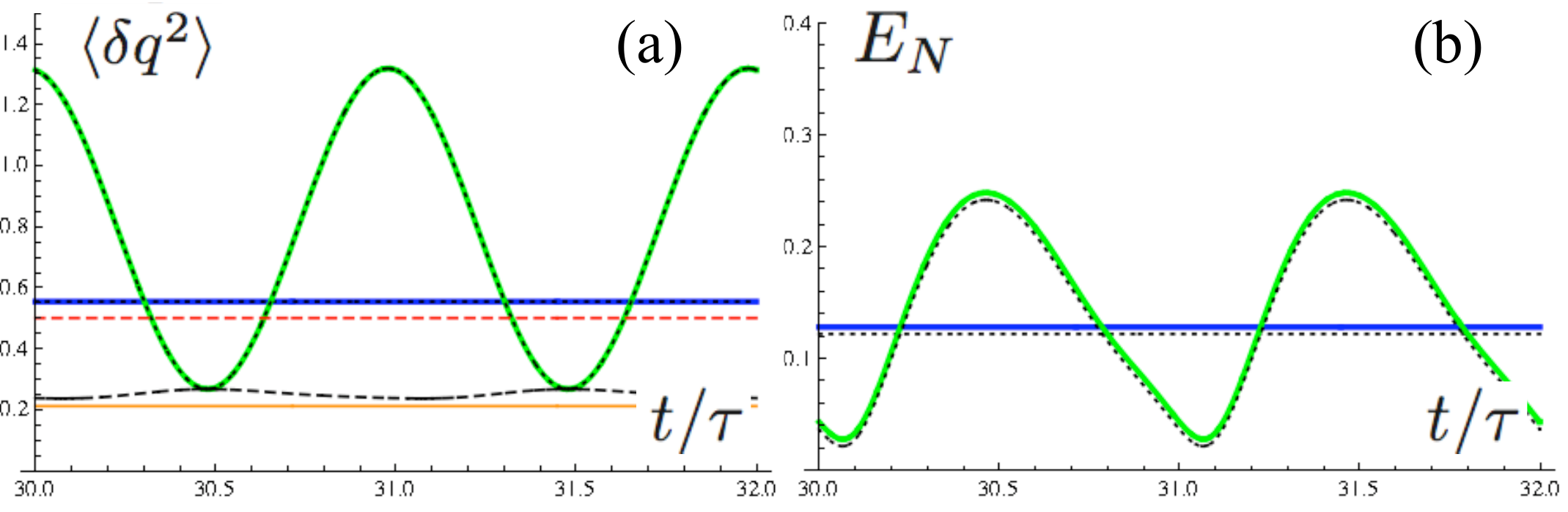}} 
\caption{(a) Variance of the mirror position and (b) light-mirror entanglement $E_N$ 
as functions of time. In both (a) and (b) the non-modulated driving regime 
(blue), the modulated driving regime (green) and the numerical
 solutions (black dashed/dotted) are plotted. (a)
 also shows the standard quantum limit 
 (red dashed) at $1/2$,
 the minimum eigenvalue of the mirror covariance 
matrix (black dashed) and its analytical estimation (\ref{varxr}) in 
the RWA (orange).} \label{varent}
\end{figure}

In order to calculate the variances of the quantum fluctuations around the classical orbits, we 
truncate the sum in Eq.\ (\ref{fulle}) to $N=2$ and we apply all the previous theory to find the covariance 
matrix $V$. In Fig.\ \ref{varent} we compare two regimes: with or without ($P_{\pm 1}=0$) modulation
(computed analytically and numerically). 
We see that the modulation of the driving field causes the emergence of significant
true quantum squeezing below the Heisenberg limit 
of the mechanical oscillator state and also the interesting phenomenon of light-mirror 
entanglement oscillations. This dynamics reminds
of the effect of parametric amplification \cite{Review,Woolley}, 
as if the spring constant of the mechanical
motion was varied in time with just twice the frequency of the mechanical motion, leading to 
the squeezing of the mechanical mode. For related ideas of reservior engineering, 
making use of bichromatic microwave coupling to a charge qubit of nano-mechanical
oscillators, see Refs.\ \cite{RablZoller}. Here, it is a more complicated joint dynamics 
of the cavity field and the mechanical mode -- where the dynamics of the first 
and the second moments can be separated -- which for large times yet
yields a similar effect. Indeed, this squeezing can directly be
measured when considering the output power spectrum, following Ref.\ \cite{NJP}, and no additional
laser light is needed for the readout, giving hence rise to a relatively simple
certification of the squeezing. Entanglement here refer to genuine quantum correlations
between the mirror and the field mode, as being quantified by the
logarithmic negativity defined as $E_N(\rho)=\log \|\rho^\Gamma\|_1$, 
essentially the trace-norm of the partial transpose, which is a proper 
entanglement measure \cite{Neg,GNeg}. 
The minimum eigenvalue of the mirror covariance 
matrix -- the logarithm thereof typically referred to as single mode
squeezing parameter 
-- is almost constant and this means that the state is always 
squeezed but that the squeezing direction continuously rotates 
in phase space with the same period of the modulation. Calling this rotating 
squeezed quadrature $\delta x_R$, a rough estimate of its variance 
can be calculated in 
the rotating-wave approximation (RWA, compare, e.g., Ref.\ \cite{RWA}),
\begin{eqnarray}\label{varxr}
 \langle\delta x_R^2\rangle=\frac{1}{2}+\bar n- \frac{2\kappa (G_0- G_{-1})(G_0 \bar n+ G_{-1}(\bar n +1))}{(\gamma_m +2\kappa)(G_0^2-G_{-1}^2+2\gamma_m \kappa)},
\end{eqnarray} 
with $\{G_n\}$ being defined as
$G(t)=\sum_{n=-\infty}^{\infty} G_n e^{i n\Omega t}$.

{\it Conclusions and outlook.} In this work we have introduced a framework of 
describing periodically amplitude-modulated optomechanical systems. Interestingly,
such a surpringly simple setting feasible with present technology \cite{Groeblacher} leads to a
setting showing high degrees of mechanical squeezing, with no feedback or
additional fields needed. We hope that such ideas contribute to
experimental studies finally certifying first quantum mechanical effects in 
macroscopic mechanical systems, constituting quite an intriguing perspective. 

{\it Acknowledgements.} This work has been supported
by the EU (MINOS, COMPAS, QAP), and the EURYI
award.

\section{Appendices}

In this appendix, we will summarize a number of 
additional statements useful for the main text.

\subsection{Differential equations}

In this appendix, we will consider several issues related to 
solutions of inhomogeneous first-order differential equations that are
used in the main text. Most of the presented material is standard
material from the theory of time-dependent periodic linear systems.
At the end, yet, we will consider a bound to the error made when assuming a 
periodic solution that is tailored to the specific situation studied above.
Consider a linear first-order system, 
\begin{equation}
	\dot x(t) = B(t) x(t),
\end{equation}
where $B(t)$ being some complex square 
matrix with entries dependent on $t\geq t_0$ with some $t_0>0$. 
Then the linear first-order system has a unique solution for $x(t_0)=x_0$
for all times $t>t_0$. The principal matrix solution is the solution
of
\begin{equation}
	\dot P(t,t_0) = B(t) P(t,t_0),
\end{equation}
with $P(t,t_0)=\id$. Floquet's theorem now states 
the following:

\begin{lemma}[Floquet's theorem] If
$B(.)$ is periodic, $B(t)=B(t+\tau)$ for some $\tau>0$ for all 
$t\geq t_0$, then the principal
matrix solution has the form
\begin{equation}
	P(t,t_0)= X(t,t_0)e^{(t-t_0)Y(t_0)},
\end{equation}	
where the matrices $X(.,.)$ and $Y(.)$ are $\tau$-periodic in all their arguments and $X(t_0,t_0)=\id$.\end{lemma}
So if one factors out an exponential,
the remainder is periodic in time. 
The eigenvalues of the
monodromy matrix
\begin{equation}
	M(t_0)= P(t_0+\tau,t_0)
\end{equation}
are then referred to as Floquet multipliers, 
and the eigenvalues of $Y(t_0)$ are known as
Floquet exponents. Floquet exponents are $t_0$ 
independent, in fact the matrix $Y(t_0)$ is similar to $Y(t_0')$ for any $t_0'$. If all Floquet exponents
have a negative real part, the system is asymptotically stable.

\begin{lemma}[Solution to inhomogeneous problem]
The solution to the inhomogeneous system 
\begin{equation}
	\dot x(t) = B(t) x(t)+g(t),
\end{equation}
with initial condition $x(t_0)=x_0$
is given by
\begin{equation} \label{inhomosol}
	x(t)= P(t,t_0)x_0 + \int_{t_0}^t
	ds  P(t,s)
	g(s).
\end{equation}
\end{lemma}

We will now show that under simple conditions that, when both $B(.)$ and $g(.)$ are $\tau$-periodic, 
we will asymptotically arrive at a solution with the same
time period. Let us define the interval
$I=[0,\tau]$.
In order to prepare that statement, we will
use the following bound. 

\begin{lemma}[Bound from Floquet exponents]\label{bf}
In the above notation, if the system is stable and if $t-t_0>1$, then
\begin{equation}\label{bound}
	\max_{u \in I}\| e^{(t-t_0+u) Y(t_0-u)}\|\leq
	c\,   n (t-t_0+\tau)^{n-1}
	e^{\lambda (t-t_0)},
\end{equation}
where 
\begin{equation}
	c=\max_{u\in I} 
	 \|W(u)\| \, \|W^{-1}(u)\|,
\end{equation}
where $W(u)$ is a similarity transformation that brings
$Y(u)$ to a Jordan normal form.
\end{lemma}

Here,
\begin{equation}
	\lambda = 
	\max_j\, \text{re} (\lambda_j),
\end{equation}
where $\lambda_j$ are the Floquet exponents.
The norm $\|.\|$ is the  norm
\begin{equation}
	\|A\|= \sup\frac{\|A x\|}{\|x\|}
\end{equation}	
induced by the usual Euclidean vector norm $\|.\|$.
So up to a constant,
the convergence is essentially governed by the largest
real part of the Floquet exponents.
To show this bound, note that
\begin{eqnarray}
	e^{(t-t_0+u) Y(t_0-u)} &=&
	W^{-1} (t_0-u)
	\left[ \oplus_j M_j (\delta)e^{\lambda_j\delta}\right] W(t_0-u),\nonumber\\
\end{eqnarray}
where
\begin{equation}
	M_j(\delta)= \left[
	\begin{array}{ccccc}
		1 & \delta& \frac{\delta^2}{2!}&\cdots & \frac{\delta^{(n_j-1)}}{(n_j-1)!}\\
		& 1 & \delta& \frac{\delta^2}{2!}& \\ 
		&  & & & \\
		&  & &\ddots &\vdots\\
		&  & & &1\\
	\end{array}
	\right],\quad \delta=t-t_0+u,
\end{equation}
and $n_j$ is the dimension of the Jordan block associated with 
the Floquet exponent  $\lambda_j$, giving rise to 
the above
expression (\ref{bound}), by 
acknowledging that 
\begin{equation}
	\|M_j(\delta)\|\leq n_j  \delta^{n_j-1}\leq n  \delta^{n-1}
\end{equation}
for all $j$, by bounding the $\|.\|$ norm from above 
by the $\|.\|_2$-norm, and since $\delta>1$. 
We can now turn back to our original problem of the periodicity of 
the asymptotic solution. 

\begin{theorem}[Asymptotic periodicity]
If the system is stable and both $B(.)$ and $g(.)$ are $\tau$-periodic (including
the case of constant $B$ or $g$), then the solution $x(.)$
of the inhomogeneous problem is asymptotically also $\tau$-periodic, 
with known 
bounds, 
in time exponentially suppressed, on the error made.
\end{theorem}

From (\ref{inhomosol}), with the new integration variable 
$u=s-\tau$, one can write
for the solution of the inhomogeneous system
\begin{eqnarray}
	x(t+\tau)&=& P(t+\tau,t_0) x_0
	 \\
	&+& \int_{t_0-\tau}^{t}
	du  P(t+\tau,u+\tau)
	g(u+\tau)\nonumber \\
	&=& P(t+\tau,t_0) x_0+ \int_{t_0-\tau}^{t}
	du  P(t,u)
	g(u) ,\nonumber
\end{eqnarray}
using the appropriate 
periodicities of $g$ and $P$. Now we try to bound the 
following quantity, which should exponentially decay to zero in order to prove the theorem,
\begin{eqnarray}\label{step1}
x(t+\tau)-x(t) &=&X(t,t_0)[e^{\tau Y(t_0)}+1]e^{(t-t_0)Y(t_0)}x_0\nonumber\\
	&+& \int_{t_0-\tau}^{t_0}du P(t,u) g(u).
\end{eqnarray}

With a new variable $v=t_0-u$, the integral in Eq.\ 
(\ref{step1}) becomes
\begin{equation}
	 \int_{0}^{\tau}dv X(t,t_0-v) e^{(t-t_0+v)Y(t_0-v)} g(t_0-v),
\end{equation}
which is bounded from above by
\begin{equation}
	\tau m \max_{v\in I} \|g(v)\|\max_{v\in I} \|e^{(t-t_0+v)Y(t_0-v)} \|,
\end{equation}
where 
\begin{equation}
	m = \max_{(t,t')\in I\times I}
	\|X(t,t')\|.
\end{equation}
Now, by using the triangular inequality, we find the following bound for the norm of the vector in (\ref{step1}),
\begin{eqnarray}
	\| x(t+\tau)-x(t)\| &\leq & m \max_{v\in I} \|e^{(t-t_0+v)Y(t_0-v)}\| \nonumber\\
 &&\times \Big[2 \| x_0 \|+\tau  \max_{v\in I} \|g(v)\|\Big]. \nonumber\\
\end{eqnarray}
Finally, by using Lemma \ref{bf}, we have 
\begin{eqnarray}\label{step2}
	\| x(t+\tau)-x(t)\| &\leq & e^{\lambda (t-t_0)} m c 
	n\, (t-t_0+\tau)^{n-1} 
	\nonumber\\
 &&\times \Big(2\| x_0 \|+\tau \max_{v\in I} \|g(v)\|\Big). 
\end{eqnarray}
Now it is clear that, in the long time limit ($t-t_0 \rightarrow \infty$), 
 the first factor $e^{\lambda (t-t_0)}$ exponentially suppresses the 
 RHS of Eq.\ (\ref{step2}) proving the theorem.

\subsection {Squeezing in the rotating wave approximation}

Suppose that we are in the particular resonance condition such that 
\begin{equation}
	\Delta(t)=\omega_m +\sum_{n\neq 0}\Delta_n e^{i n \Omega t}
\end{equation}	
and $\Omega=2 \omega_m$. Then, in the limit of
 $\omega_m\gg
 |G(t)|,\kappa$, a simple analytical expression for the covariance 
 matrix can be obtained in the rotating wave approximation.
We observe that we have two arbitrary degrees of freedom that we can 
choose to simplify the calculation: The global phase of the driving laser 
and the initial time corresponding to $t=0$.  If we expand the coupling coefficient $G(t)=
\sum_{n=-\infty}^\infty
G_n e^{i n \Omega t}$, then we can assume without loss of generality 
assume that 
$G_0,G_{-1}\in \mathbb R$.

Now we move to an interaction picture introducing the slowly varying bosonic operators, 
\begin{equation}
	a_s=a_s e^{i\omega_m t},\
	b_s=b e^{i\omega_m t}.
\end{equation}
In this reference frame, if we neglect terms rotating at frequency $2 \omega_m$, 
we obtain an equation analogous to Eq.\ (\ref{secdiff}):
\begin{equation}\label{secdiff2}
 \dot{V_s}(t)=A_s V_s + V_s A_s^T + D_s
\end{equation}
where, 
\begin{equation}\label{As}
  A_s=\frac{1}{2} \left[\begin{array}{cccc}
    	-\gamma_m 	 & 0	& 0 		& G_{-1}-G_0 \\
     	0 & -\gamma_m	&G_{-1}+ G_0	& 0  \\
        0 & 	G_{-1}-G_0	& -2\kappa 	&0) \\
    	G_{-1}+G_0  &     0	& 0 	& -2\kappa
  	\end{array}\right],
\end{equation}
and 
\begin{equation}	
	D=\text{diag}
	\bigl(\gamma (\bar n+{1}/{2}),\gamma(\bar n+{1}/{2}),\kappa,\kappa\bigr).
\end{equation}
We observe that, in the RWA, only the coefficients $G_0$ and $G_1$ matter. They correspond to the cooling and the heating sidebands of the input driving laser.

The stability condition for the differential Eq.\ 
(\ref{secdiff2}) is $G_{-1}^2- G_0^2 \le 2 \gamma_m \kappa$, 
which is always satisfied if the cooling process is predominant with respect to the heating. 
Differently from Eqs.\ (\ref{secdiff}), Eq.\ 
(\ref{secdiff2}) has constant coefficients 
and therefore, in the stable regime, $V_s(t)$ reaches 
the an asymptotic constant value $V_s=\lim_{t\rightarrow \infty} V_s(t)$.

The matrix $V_s$ can be calculated by imposing the derivative in (\ref{secdiff2}) equal to zero and solving the remaining linear system. We report only the mirror variances: 
$(V_s)_{1,1}=f_-$, $(V_s)_{2,2}=f_+$ and $(V_s)_{1,2}=0$, where
\begin{equation}\label{f+-}
 f_{\pm}=\frac{1}{2}+\bar n- \frac{2\kappa (G_0 \pm G_{-1})[G_0 \bar n\mp G_{-1}(\bar n +1)]}{(\gamma_m +2\kappa)(G_0^2-G_{-1}^2+2\gamma_m \kappa)}.
\end{equation}
Three particular limits. If $G_{-1}=0$, which corresponds to a red 
detuned driving laser without modulation, we observe the usual cooling of the mirror. 
For $G_{-1}=G_0$, we  recover the QND measurement setting 
studied in Ref.\ \cite{Clerk}, where a symmetric driving with 
opposite detuning couples the light only with a quadrature of the mirror. Indeed, 
the variance $V_{s11}$ is unaffected by the driving while the other is 
increased due to radiation pressure noise. Finally if $G_{-1}<G_0$, 
we observe that the mirror can reach a steady state which is both 
cooled and squeezed without the need of any feedback and control.

\subsection {Squeezed environments versus modulation}

In recent work \cite{Jaehne} it has been shown that it is possible to squeeze 
the mechanical mode by using a constant driving field with $\Delta=\omega_m$ 
and an additional squeezed vacuum beam with frequency resonant with cavity $\omega_s=\omega_c$. 
In the RWA formulation used in the previous section, this corresponds to a coefficient matrix $A'$ having the same structure as in Eq. (\ref{As}) but with $G_0=G'$ and $G_{-1}=0$, while 
\begin{equation}
	D'=\text{diag}
	\bigl(\gamma (\bar n+{1}/{2}),\gamma(\bar n+{1}/{2}), s \kappa,s^{-1}\kappa\bigr), 
\end{equation}
where $s$ is a parameter quantifying the squeezing of the environment. If we solve the system for the mirror variances, instead of Eq. (\ref{f+-}) we get
\begin{equation}\label{f2}
	 f'_{\pm}=\frac{1}{2}+\bar n-\frac{\kappa G^2 (2 \bar n+1-s^{\pm 1}) }{
	 (\gamma_m + 2 \kappa) (G^2 + 2 \gamma_m \kappa)}.
\end{equation}
Now we show that, within the validity of the RWA, for what concerns 
the squeezing of the mirror, this approach is equivalent to our scheme in 
which the cavity is driven by a single amplitude modulated laser and no squeezing
is required.
In fact, by choosing the modulation parameters such that 
\begin{eqnarray}
	G_0&=&G(s^{-{1}/{2}}+s^{{1}/{2}})/2,\\
	G_1&=&G(s^{-{1}/{2}}-s^{{1}/{2}})/2,
\end{eqnarray}	 
the previous formula given in (\ref{f+-}) reduces exactly to Eq.\ (\ref{f2}). 
This means that the two approaches are formally equivalent as far as the
squeezing is concerned and the 
choice between one or the other depends only on technical factors.

Finally, we may ask if a good mechanical squeezing is possible by 
just driving the cavity with a single squeezed non-modulated field with
 $\Delta=\omega_m$.  Differently from the proposal of Ref.\ \cite{Jaehne}, in this 
 case the squeezed input noise is not resonant with the cavity mode but rotates 
 with a frequency $\omega_m$. This means that, in the RWA, the squeezing is 
 averaged to zero and the optical environment looks like an effective thermal
  bath. Therefore, the only convenient choice is to use an additional squeezed 
  beam which is resonant with cavity as in Ref.\ \cite{Jaehne}. Only in 
  this way the squeezing direction of the environment does not rotate and 
  can have a significant effect on the system dynamics.


\begin{thebibliography}{100}
\bibitem{Pedantic}
	Pedantically speaking -- unless one is in the limit of
	very weak coupling compared to the free evolution (``quantum optical
	Markovian limit'') -- this requires an Ohmic spectral density of the
	heat bath and a high temperature (``quantum Brownian motion
	limit''), see, e.g.\ Ref.\ \cite{Breuer}. For the considered system,
	the Markovian approximation is typically a very good one.
\bibitem{GNeg}
	$E_N(\rho) = - \sum_{i=1}^2\min(0, \log(c_i))$, where $c_{1,2}$ 
	are the eigenvalues of $2|V^{1/2} iS V^{1/2}|$, where $S$ is the $4\times4$-matrix with
	$S_{1,2}=S_{4,3}=1$, $S_{2,1}=S_{3,4}=-1$ and zero otherwise.
\end{thebibliography}
\end{document}